\documentclass[aip,apl,reprint,amsmath,amssymb,amsfonts, floatfix, showpacs,intlimits]{revtex4-1}

\usepackage{graphicx}

\usepackage[usenames, dvipsnames]{color}
\definecolor{Red}{rgb}{1.00, 0.00, 0.00}

\newcommand{\iext}{i_{\mathrm{ext}}(t)}
\newcommand{\rstno}{r_{\mathrm{stno}}(t)}
\newcommand{\vstno}{v_{\mathrm{stno}}(t)}
\newcommand{\vlpf}{v_{\mathrm{lpf}}(t)}
\newcommand{\vspec}{v_{\mathrm{spec}}(t)}

\newcommand{\idc}{I_\mathrm{DC}(t)}

\begin{document}

	\title{Ultra-fast wide band spectrum analyzer based on a rapidly tuned spin-torque nano-oscillator}
	\author{Steven Louis}
	\email[]{slouis@oakland.edu}
	\affiliation{Department of Electrical and Computer Engineering, Oakland University, 2200 N. Squirrel Rd., Rochester, MI, 48309, USA}
	
	\author{Olga Sulymenko}
	\affiliation{Faculty of Radio Physics, Electronics and Computer Systems, Taras Shevchenko National University of Kyiv, 64/13 Volodymyrska str., Kyiv 01601, Ukraine}
	
	\author{Vasyl Tyberkevych}
	\affiliation{Department of Physics, Oakland University, 2200 N. Squirrel Rd., Rochester, MI, 48309, USA\looseness=-1}
	
	\author{Jia Li}
	\affiliation{Department of Electrical and Computer Engineering, Oakland University, 2200 N. Squirrel Rd., Rochester, MI, 48309, USA}			

	\author{Daniel Aloi}
	\affiliation{Department of Electrical and Computer Engineering, Oakland University, 2200 N. Squirrel Rd., Rochester, MI, 48309, USA}	
	
	\author{Oleksandr Prokopenko}
	\affiliation{Faculty of Radio Physics, Electronics and Computer Systems, Taras Shevchenko National University of Kyiv, 64/13 Volodymyrska str., Kyiv 01601, Ukraine}
	
	\author{Ilya Krivorotov}
	\affiliation{Department of Physics and Astronomy, University of California, Irvine, California, 92697, USA\looseness=-1}
	
	\author{Elena Bankowski}
	\affiliation{U.S. Army TARDEC, Warren, MI, 48397, USA}
	
	\author{Thomas Meitzler}
	\affiliation{U.S. Army TARDEC, Warren, MI, 48397, USA}
	
	\author{Andrei Slavin}
	\affiliation{Department of Physics, Oakland University, 2200 N. Squirrel Rd., Rochester, MI, 48309, USA\looseness=-1}
	
	\date{\today}

\begin{abstract}
A spintronic method of ultra-fast broadband microwave spectrum analysis is proposed. It uses a rapidly tuned spin torque nano-oscillator (STNO), and does not require injection locking. This method treats an STNO generating a microwave signal as an element with an oscillating resistance. When an external signal is applied to this ``resistor'' for analysis, it is mixed with the signal generated by the STNO. The resulting mixed voltage contains the ``sum'' and ``difference'' frequencies, and the latter produces a DC component when the external frequency matches the frequency generated by the STNO. The mixed voltage is processed using a low pass filter to exclude the ``sum'' frequency components, and a matched filter to exclude the dependence of the resultant DC voltage on the phase difference between the two signals. It is found analytically and by numerical simulation, that the proposed spectrum analyzer has a frequency resolution at a theoretical limit in a real-time scanning bandwidth of 10~GHz, and a frequency scanning rate above 1~GHz/ns, while remaining sensitive to signal power as low as the Johnson-Nyquist thermal noise floor.

\end{abstract}

\maketitle

Spectrum analyzers are critically important instruments with applications in engineering, science, and medicine\cite{Boashash2015book, Rangayyan2015book}. Historically, spectrum analyzers have been implemented with either swept-tuned or Fourier methods. More recently, real-time spectrum analyzers have started to use a combination of these methods and vector signal analysis. Despite substantial technological improvements, current real time spectrum analyzers for demanding applications, such as pulsed radar frequency determination or electronic signal intelligence, are exceedingly complex and/or computationally expensive\cite{Boashash2015book}.

We propose to use a rapidly tuned spin torque nano-oscillator (STNO) \cite{kiselev2003microwave,rippard2005injection,chen2016spin,deac2008bias,slavin2009nonlinear} based on a magnetic tunnel junction (MTJ) to perform fast, broadband spectrum analysis with frequency scanning rates and bandwidths that exceed current state of the art, all while remaining sensitive to signals with power levels as low as the Johnson-Nyquist thermal noise floor.
STNOs are nano-sized low power microwave auto-oscillators that can be tuned over a wide frequency range by adjusting a driving bias DC current. They have a number of interesting features, including low operating power, compatibility with CMOS technology, nonlinear synchronization behavior, operation from below 1 GHz to above 65 GHz, high modulation rates, and the possibility of a radiation-hard construction \cite{slavin2009nonlinear,louis2017low,zhou2010oscillatory,moriyama2012phase,kaka2005mutual,urazhdin2010fractional,villard2010ghz,zeng2012high,slavin2005nonlinear, bonetti2009spin,quinsat2014modulation,zhou2008spin,pufall2005frequency}. At low frequencies ($f<3$ GHz), they have been constructed to operate in the absence of a bias magnetic field\cite{zeng2013ultralow}. STNO oscillations occur in an MTJ when a DC electric current of sufficient amplitude excites the free layer magnetization to precess with a microwave frequency due to the spin-transfer torque effect\cite{slonczewski1996current,berger1996emission}. These magnetization oscillations can be detected macroscopically through the effect of tunneling magnetoresistance (TMR). For the purposes of this paper, an STNO (a current-driven MTJ with an oscillating TMR) will be treated as an oscillating resistor.

\begin{figure}
\centering
\includegraphics{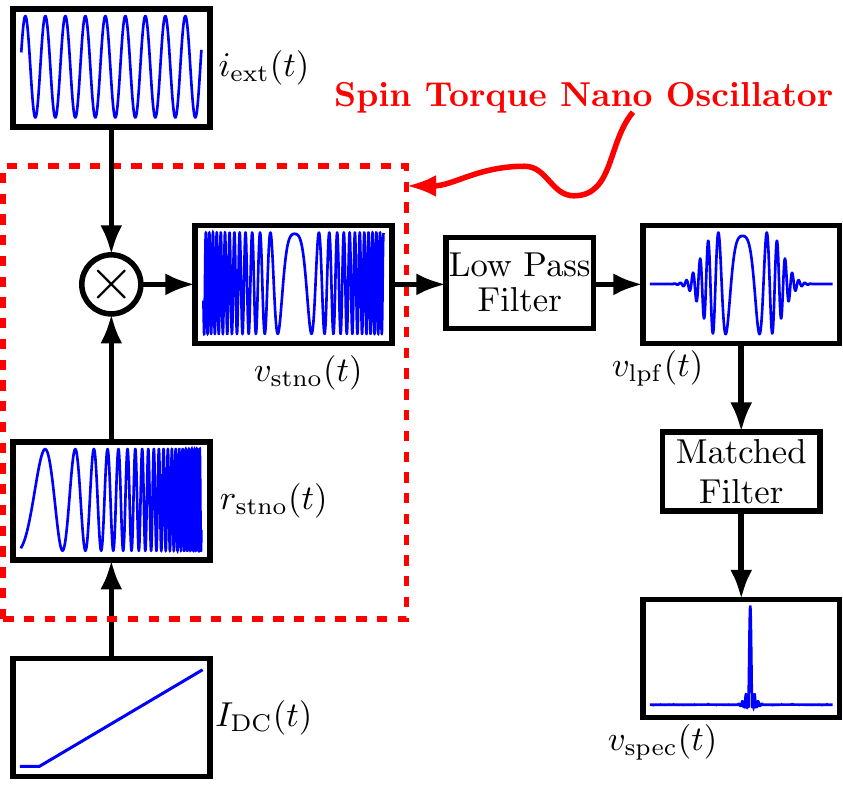}
\caption{Block diagram of a spintronic spectrum analyzer. The MTJ based STNO tunneling magneto-resistance $\rstno$, which is driven by ramped current $I_\mathrm{DC}$, is multiplied by external microwave current $\iext$ to produce STNO voltage $\vstno$. After passing through a low pass filter and a matched filter, a spectrum $\vspec$ is produced.}
\label{schematic}\end{figure}

The configuration of a proposed spectrum analyzer is introduced by a block diagram in Fig.~\ref{schematic}. In this diagram, a ramped bias DC current $\idc$ drives the STNO (outlined by a red dashed line) which generates a signal with frequency $f(t)$ that linearly increases with time. The TMR of the STNO, $\rstno$, oscillates with the same frequency $f(t)$. An external microwave signal, which will be analyzed, is represented by a current $\iext=I_\mathrm{ext}\cos(2\pi f_\mathrm{ext} t)$, where $I_\mathrm{ext}$ is the external signal amplitude and $f_\mathrm{ext}$ is the external signal frequency. The external signal is analyzed in three steps. First, $\iext$ and $\rstno$ are combined via Ohm's law to produce voltage $\vstno$ which has a high frequency component ($ f(t)+f_\mathrm{ext} $) and low frequency component ($ f(t)-f_\mathrm{ext} $). Second, a low pass filter with a cutoff frequency $f_\mathrm{c}$ is used to eliminate the high frequency component and a portion of the low frequency component. Third, a matched filter removes the dependence on the phase difference between the mixed signals and improves the signal-to-noise ratio (SNR). The matched filter output, $\vspec$, provides the spectrum of $\iext$.

Overall, the method of operation of the proposed spectrum analyzer is similar to a traditional swept-tuned spectrum analyzer, with the exception that frequency tuning is performed entirely by a single STNO. Using an STNO is advantageous, because inside a frequency range as wide as 10 GHz, it can be tuned rather fast ($>5$GHz/ns) due to its small size, and hence, inherently low capacitance and inductance\cite{louis2016,manfrini2009agility}.

This work differs from a previous study, which performed spectrum analysis using the injection locking properties of an STNO sweeping a wide bandwidth\cite{louis2017low}. The previous work showed that a minimum time and energy was required for an STNO to injection lock to an external signal, thus precluding the use at faster scan rates for low power signals. In contrast, for an STNO treated as a tunable oscillating resistor, as in the present study, \emph{no minimum threshold energy is required} for the signal detection, and, thus, only the noise floor determines the minimum detectable signal regardless of the scan rate. We will show that spectrum analysis with an STNO and a matched filter can be performed with a speed which is limited by the supporting electronics.

This relatively simple scheme has several advantages, including: \textit{i}) wide scanning bandwidth, \textit{ii}) high scanning speed, \textit{iii}) high sensitivity to low power signals, \textit{iv}) invariance to phase shifts, and \textit{v}) frequency resolution at a theoretical limit. We will demonstrate each of these properties by numerical simulations.

The magnetization dynamics in the MTJ free layer under the action of a DC current can be modeled using the Landau-Lifshitz-Gilbert-Slonczewski equation\cite{slavin2009nonlinear}:\begin{equation}	 \frac{\mathrm{d}\mathbf{m}}{\mathrm{d}t} = |\gamma| \mathbf{B}_{\rm eff}\times\mathbf{m}+\alpha_{\rm G}\mathbf{m}\times\frac{\mathrm{d}\mathbf{m}}{\mathrm{d}t}+|\gamma|\alpha_{\rm J}\idc\mathbf{m}\times[\mathbf{m}\times\mathbf{p}]\label{llgs}.\end{equation}
In this equation, $\mathbf{m}$ is the normalized unit-length magnetization of the free layer in the macrospin approximation, $\gamma=-2\pi28~{\rm GHz/T}$ is the gyromagnetic ratio, $\mathbf{B}_{\rm eff}=\mathbf{B}_{\rm ext}-\mu_0 M_{\rm s} (\mathbf{m}\cdot\mathbf{\hat{z}})\mathbf{\hat{z}}$ is the effective field, $|B_{\rm ext}|=1.5~{\rm T}$ is the external field that is applied perpendicular to the free layer plane, in the $\mathbf{\hat{z}}$ direction. The free layer saturation magnetization is $\mu_0 M_s=0.8~{\rm T}$, $\idc$ is the bias DC current that controls the STNO frequency, $\alpha_{\rm G}=0.01$ is the Gilbert damping constant, and $\alpha_{\rm J}=\hbar \eta_0 / (2 \mu_0M_{\rm s}eV)$ is the spin-torque coefficient, where $\hbar$ is the reduced Planck constant, $\eta_0=0.35$ is the spin polarization efficiency, $\mu_0$ is the free space permeability, $e$ is the fundamental electric charge, and $V=3\times 10^4~{\rm nm}^3$ is the volume of the free layer. The direction of the spin current polarization was chosen as $\mathbf{p}=\cos(\beta)\mathbf{\hat{x}}+\sin(\beta)\mathbf{\hat{z}}$ with $\beta=30^\circ$. In this configuration, the threshold current of microwave signal generation in the STNO was $I_\mathrm{th}=2.32~{\rm mA}$. The STNO magnetization can be found by solving (\ref{llgs}) numerically, thus allowing computation of the resistance $\rstno=R_\mathrm{stno} (\mathbf{m}\cdot\mathbf{p})$, where $R_\mathrm{stno}=1$~k$\Omega$ is the average STNO resistance. All parameters were chosen to have typical values, that were used in previous publications\cite{louis2017low,zhou2010oscillatory}.

\begin{figure}
\centering
\includegraphics{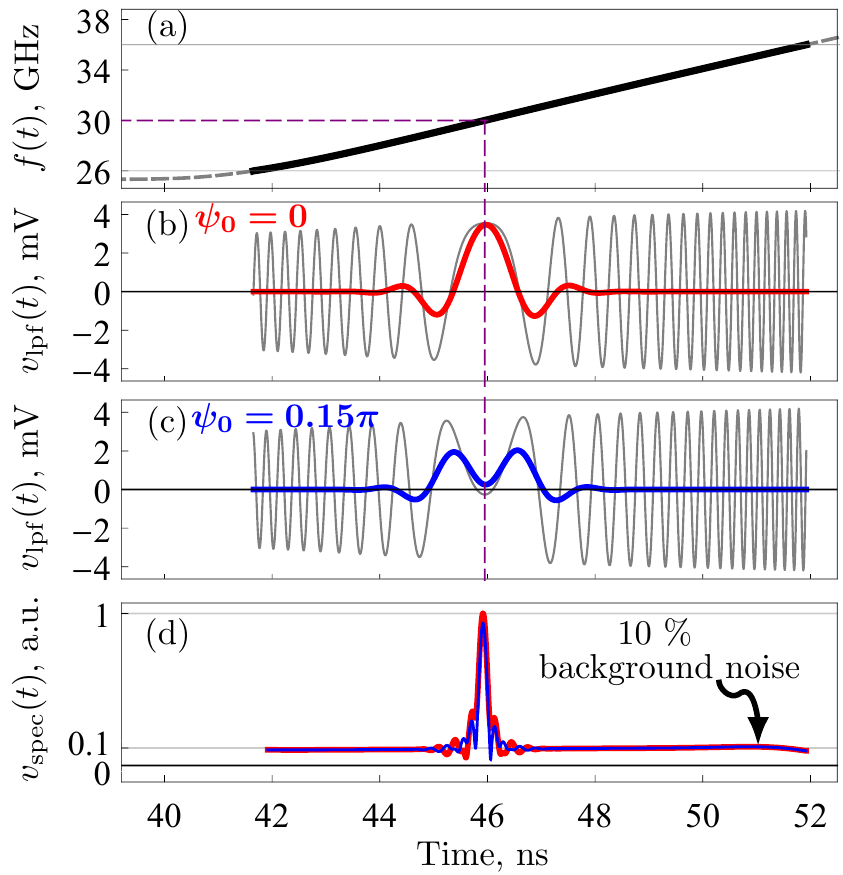}
\caption{(a) The frequency of STNO, $f(t)$, in response to $\idc$. The gray dashed line shows the STNO frequency, which is constant at 25.4 GHz until 40 ns, then increases in response to a ramped DC current. The solid black line shows the STNO scanning from 26 to 36 GHz. (b) STNO output with $\psi_0=0$. Grey line shows $\vlpf$ with $f_c=15$ GHz, and thick red line shows $\vlpf$ with $f_c=2$ GHz. (c) STNO output with $\psi_0=0.15\pi$. (d) Matched output filter, $\vspec$, for two phases, with red for $\psi_0=0$ and blue for $\psi_0=0.15\pi$. Simulated with $f_\mathrm{c}=4.5$ GHz and $f_\mathrm{m}=32$ GHz.}
\label{theory}
\end{figure}

When the STNO is driven by a current exceeding $I_\mathrm{th}$, the magnetization $\mathbf{m}$ begins a stable precession about $\mathbf{B}_{\rm eff}$. This precession will lead to the sustained oscillations of the STNO resistance $\rstno$. The frequency of $\rstno$, as acquired by simulation, are shown by a gray dashed line in Fig. \ref{theory}(a). In the first 40 ns of this simulation, the bias current was held constant with $\idc=3.0$ mA, allowing $\rstno$ to oscillate steadily at a frequency of 25.4 GHz. After 40 ns, $\idc$ was increased with a slope of $\approx 1.2$ mA/ns. This current increase caused the STNO oscillation frequency to increase at the constant scan rate of $\rho=1$ GHz/ns, with a frequency: \begin{equation}f(t)= f_0 + \rho t,\label{freq}\end{equation} where $f_0$ is an initial frequency. The thick black line in Fig. \ref{theory}(a) shows the interval where spectrum analysis was performed, when $f(t)$ is between 26 and 36 GHz.

All simulations reported in this Letter were performed without noise, and with an external signal with a power of $\approx 0.05$ pW ($I_\mathrm{ext}=10$ nA). The power level was chosen to show that this method of spectrum analysis will work for any power level above the Johnson-Nyquist noise floor.

When the STNO generates a signal of a variable frequency (\ref{freq}), the STNO resistance is described by the expression $\rstno=R_\mathrm{stno}\cos\big(2\pi f_0 t + \pi \rho t^2 + \psi_0\big)$, where the constant of integration $\psi_0$ physically represents the phase difference between $\iext$ and $\rstno$.

When an external microwave signal $\iext$ is introduced for analysis, it passes through the MTJ, producing a voltage $\vstno=\iext\rstno$. As this voltage is the product of two sinusoidal signals, it will have a high frequency component ($ f(t) +f_\mathrm{ext} $) and a low frequency component ($ f(t)-f_\mathrm{ext} $). A low pass filter is then applied to $\vstno$ to remove frequencies above $f_\mathrm{c}$, and  the low frequency voltage can be expressed as:
\begin{equation}\vlpf=I_\mathrm{ext}R_\mathrm{stno}\cos\big(\phi(t,f_\mathrm{ext}) + \psi_0\big).\label{vsteq}\end{equation}
Where $\phi(t,f_\mathrm{ext})=2\pi (f_0-f_\mathrm{ext}) t + \pi \rho t^2 $ is the phase of the low frequency component of $\vstno$.


It is important to note, that experimentally the phase difference $\psi_0$ cannot be controlled, so this numerical simulation was performed  with two representative phases, $\psi_0=0$ and $\psi_0=0.15\pi$. The results of these two simulations are shown in Fig. 2(b) and 2(c), with a thin gray line representing $\vlpf$ with $f_\mathrm{c} = 15 $ GHz. This line in both Fig. 2(b) and 2(c) shows that the frequency of the voltage is at a minimum near $t=46$ ns. This low frequency voltage coincides in time with the moment when the frequencies of $\rstno$ and $\iext$ are the same. This is emphasized by the thick red and blue lines, which show $\vlpf$ with $f_\mathrm{c} = 2 $ GHz. Both of these lines show increased amplitude when the frequencies of the two signals are the same, and, indeed, the red line in Fig. 2(b) can be used to precisely determine the time when the two frequencies are equal. However, this is not possible when $\psi_0=0.15\pi$, as the dual peaks seen in the blue line of Fig. 2(c) could be produced instead by two external signals.

A matched filter can be used to limit the influence of $\psi_0$ and vastly improve the SNR. Matched filters operate by producing a strong peak when an input signal matches a template. Noting that (\ref{vsteq}) can be represented as $\propto e^{i\phi(t,f_\mathrm{ext})}e^{i\psi_0}+e^{-i\phi(t,f_\mathrm{ext})}e^{-i\psi_0}$, we choose a template for the matched filter as $h(t)=e^{-i\phi(t,f_\mathrm{m})}$, where $f_\mathrm{m}$ is an arbitrary frequency in the interval of the spectrum analysis. Choosing this $h(t)$ maximizes the SNR, and, hence, the precision of the frequency determination.

Applying the above described matched filter to $\vlpf$ produces the following output spectrum: \begin{equation}\vspec=h(t)*\vlpf\label{spec}\end{equation} where `$*$' is the symbol for convolution. Fig. \ref{theory}(d) shows $\vspec$ for signals with phases $\psi_0=0$ and $\psi_0=0.15\pi$. It is evident that both curves show a sharp peak at $f(t\approx46 ~\mathrm{ns})=30$ GHz, and that both peaks are independent of the phase difference $\psi_0$. Both curves also show a relatively flat background noise of about 10\% the signal maximum, and a minor phase dependance in the neighborhood of the peak. The curves in Fig. \ref{theory}(d) represent the primary result of this Letter; they show that with (\ref{spec}), an STNO is theoretically capable of detecting a 0.05 pW signal while scanning a 10 GHz interval at a rate of 1 GHz/ns, and that the detection is independent of any variation of the phase difference $\psi_0$.

How this particular matched filter works can be seen by filtering each of the exponential terms in (\ref{vsteq}) independently. When the positive exponent is filtered by $h(t)$: \begin{equation}h(t)*e^{i\phi(t,f_\mathrm{ext})}e^{i\psi_0} =\tfrac{1}{2\pi\rho}\delta(t-t_\mathrm{ext})
,
\label{pos}
\end{equation} the output signal has no phase dependence, and a distinct peak is observed at the time $t_\mathrm{ext}=(f_\mathrm{m}+f_\mathrm{o}-f_\mathrm{ext})/\rho$, as shown in Fig. \ref{theory}(d).
When the negative exponent of (\ref{vsteq}) is filtered by $h(t)$:
\begin{equation}h(t)*e^{-i\phi(t,f_\mathrm{ext})}e^{-i\psi_0}\label{neg}=\tfrac{1}{2\sqrt{\pi\rho}}\end{equation} the result is a constant.

Equations (\ref{pos}) and (\ref{neg}) represent an ideal case, where $\rstno$ and $\iext$ are non-zero for all time, $-\infty<t<\infty$. For more realistic signals that exist during a finite interval, as in Fig. \ref{theory}, the peak in (\ref{pos}) will be broadened, while (\ref{neg}) will have a weak phase dependence near $t_\mathrm{ext}$, and the peak will have a phase dependent amplitude. It is notable, that for smaller values of $\rho$, indicating slower scan rates, the background influence from (\ref{neg}) will be reduced, improving the SNR.

\begin{figure}
\centering
\includegraphics{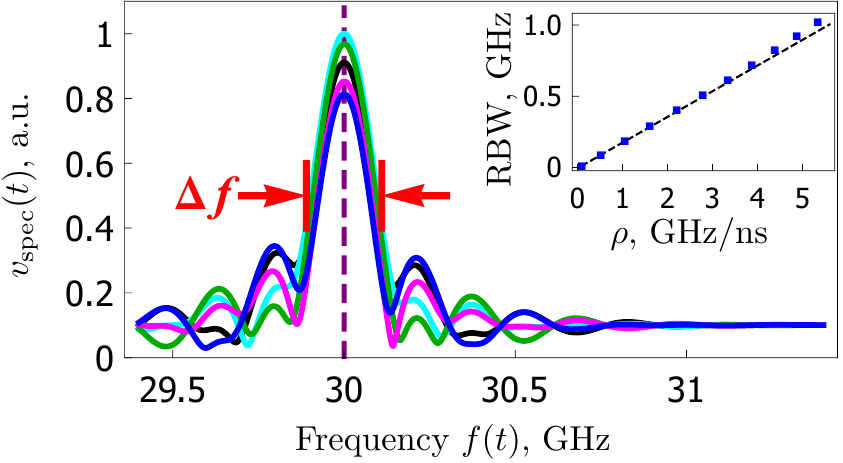}
\caption{Phase invariance, external signal amplitude, and RBW. Five STNO scans with $f_\mathrm{ext}=30$ GHz, $\rho=1$ GHz/ns and varied $\psi_0$. The peak, independent of $\psi_0$, occurs at 30 GHz, with $\Delta f = 220$ MHz. The peak amplitude has an uncertainty of about 8\%. (inset) Comparison of RBW and $\rho$.}
\label{res}
\end{figure}

To further demonstrate that $\vspec$ is independent of $\psi_0$, the simulation above was repeated with 50 different phases, ranging from $\psi_0=0$ to $\pi$ with a 0.02$\pi$ step. Five typical peaks are shown in the Fig. \ref{res}. The mean peak frequency for all 50 simulations was found to be 30.001 GHz with an error of just 0.001\%. This shows that regardless of the phase difference between the external signal and the STNO, the frequency can be determined with a sufficiently high precision.

In spectrum analysis, the frequency resolution, known as the resolution bandwidth (RBW), is a measure of the minimum separation required to distinguish two neighboring frequencies, with a low RBW preferred. The RBW of this system, as simulated, is approximately equal to the 3 dB linewidth of a peak, which in Fig. 3 is $\Delta f \approx 220$ MHz. An estimate for a theoretical limit for the RBW is $\Delta f_\mathrm{res}\approx\tfrac{\rho}{f_\mathrm{c}}$, where $f_\mathrm{c}=4.5$ GHz is the bandwidth of the low pass filter used to simulate Fig. 2(d) and Fig. 3.
A comparison of the simulated and theoretical RBW for a variety of scan rates is presented in the inset of Fig. \ref{res}, with blue squares denoting $\Delta f$ and the dashed black line indicating $\Delta f_\mathrm{res}$. It is evident, that the RBW for different scan rates remains near the theoretical limit for scan rates as high as 5 GHz/ns.

In addition to the frequency determination, the amplitude of the external signal, $I_\mathrm{ext}$, can be determined from the amplitude of the detection peak. However, there is a minor variation in peak amplitude that is dependent on $\psi_0$. For $\rho=1$ GHz/ns, the uncertainty is about 8\%. This uncertainty can be reduced by further signal processing.


\begin{figure}
\centering
\includegraphics{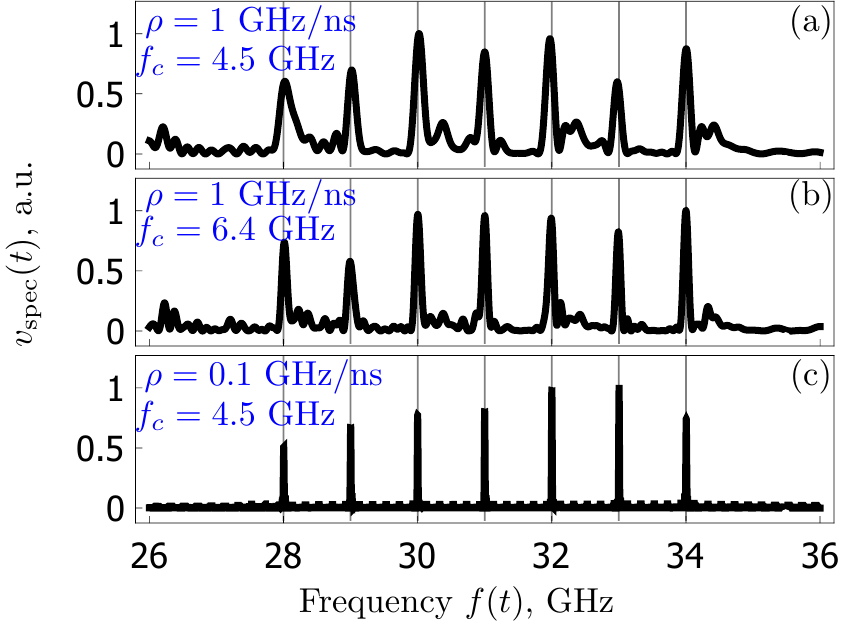}
\caption{Spectra produced by identical 10 nA signal at every integer from 27 to 35 GHz and a random phase. Responses from the following scan parameters: (a) $\rho=1$ GHz/ns and $f_c=4.5$ GHz (b) $\rho=1$ GHz/ns and $f_c=6.4$ GHz (c) $\rho=0.1$ GHz/ns and $f_c=4.5$ GHz.}
\label{example}
\end{figure}

In principle, this ``spintronic'' design for microwave spectrum analysis can be readily implemented, because all the constituent components have already been used in an experimental settings. MTJs have been extensively manufactured for many years, and STNOs have been studied for over a decade. Similarly, matched filters were developed for radar applications in the 1940s, and a matched filter like (\ref{spec}) can be implemented digitally. Digitization of $\vstno$ requires an anti-aliasing filter, which motivates our choice of $f_\mathrm{c}=4.5$ GHz, which is lower than currently available digitization rates of 6.4 GSPS\cite{ti}.

The last figure demonstrates the analysis of a signal having a more complicated spectrum. Fig. 4 shows three simulations where an STNO-based spectrum analyzer operated on an external signal that had frequencies at every integer frequency from 28 GHz to 35 GHz, each with a distinct random phase $\psi_0$ between 0 and $2\pi$. The external currents used for all three simulations were identical. Fig. 4(a) shows the response with $\rho=1$ GHz/ns and $f_c= 4.5$ GHz. With these characteristics, all peaks are correctly determined despite a rising level of a background noise. Fig. 4(b) was simulated with the same scan rate $\rho = 1$ GHz/ns and a higher cutoff frequency of $f_c= 6.4$ GHz. It is evident that the higher $f_c$ value decreases the linewidth and the background noise level. This suggests, that as the digitization rates improve, the higher scan rates with lower RBW will be possible. The spectral quality is further improved in Fig. 4(c), which uses a slower scan rate of $\rho=100$ MHz/ns, and $f_c=4.5$ GHz. It is evident, that at slower scan rates, as expected, the RBW improves, and the background noise is substantially reduced. The variation in amplitude with respect to frequency in these detected signals is expected, and is caused by the change in the angle of precession in $\mathbf{m}$, and could be easily normalized using standard methods of digital signal processing.

In conclusion, an STNO with a matched filter is capable of performing spectrum analysis at high speeds, high sensitivity, and high resolution. 
It is able to determine the frequency of an external signal with a high precision.
 The performance of this system relies on a matched filter to remove the phase-related distortion. The sensitivity of the spectrum analyzer was found to be at the the level of the Johnson-Nyquist noise floor, and the RBW of the system was found to be near the theoretical limit.

This work has been supported by the Grants Nos. EFMA-1641989 and ECCS-1708982 from the NSF of the USA, and by the DARPA M3IC Grant under the Contract No. W911-17-C-0031. It contains the results of studies conducted by President's of Ukraine grant for competitive projects (F 78) and by grant F 76 from the SFFR of Ukraine. This work was also supported in part by the grants 16BF052-01 and 18BF052-01M from KNU and grant 7F from NASU.

\bibliography{STNORampBib}

\end{document}